# Inverse Orbital Hall Effect Discovered from Light-Induced Terahertz Emission

Yong Xu[1,2], Fan Zhang[1,2], Yongshan Liu[1,2], Renyou Xu[1,2], Yuhao Jiang[1], Houyi Cheng[1,2], Albert Fert[1,3*], and Weisheng Zhao[1,2*]

[1]MIIT Key Laboratory of Spintronics, School of Integrated Circuit Science and Engineering, Beihang University, Beijing 100191, China

[2]Hefei Innovation Research Institute, Beihang University, Hefei 230013, China

[3]Unité Mixte de Physique, CNRS, Thales, Université Paris-Saclay, Palaiseau 91767, France

These authors contributed equally: Yong Xu, Fan Zhang

*Corresponding author: albert.fert@cnrs.fr, weisheng.zhao@buaa.edu.cn

**Abstract:** Recent progress in orbitronics reveals the possibility of using orbit current as an information carrier. The interconversion between orbit currents and charge currents is crucial for orbit information processing. Although orbit currents can be created from charge currents via the orbital Hall effect, the conversion from orbit currents into charge currents has been observed only in very few systems due to the lack of a reliable orbit current source and the disturbance of the omnipresent inverse spin Hall effect. In this study, we show that ultrafast pulses of orbit current can be generated in Ni layers by femtosecond laser pulses. We demonstrate that, by injecting such orbit current pulses into nonmagnetic metals, a transient charge current is induced and emits terahertz electromagnetic pulses. The nonmagnetic metal layer acts as a converter of the orbit current into the charge current. The discovery of the generation and conversion of light-induced orbit current opens a new route for developing future orbitronic devices.

The conversion from a charge current $j_C$ to a spin current $j_S$ has been studied extensively in systems with strong spin-orbit coupling [1,2]. The most successful example of the conversion mechanism is the spin Hall effect (SHE) for nonmagnetic metals containing heavy atoms [3–6]. For two-dimensional electron gases at Rashba interfaces and the surfaces or interfaces of topological insulators, the charge-to-spin conversion is named the Rashba-Edelstein effect (SREE, with S for spin in our notation) [7–10]. In ferromagnet (FM)/nonmagnet (NM) heterostructures, the spin currents induced by SHE or SREE are strong enough to reverse the magnetization of the FM material by Spin Transfer Torque [11,12]. Therefore, both SHE and SREE have attracted much attention due to their technological significance in developing future magnetic memory devices. Their inverse effects, namely, inverse SHE (ISHE) [13] and inverse SREE (ISREE) [14], convert a spin current $j_S$ into a charge current $j_C$. These effects have been widely utilized for detecting spin currents generated by other stimuli, such as the heat current [15], the charge current [5,6], and the spin-pumping technique [16–18]. In recent years, ISHE and ISREE have also been utilized to generate ultrafast charge pulses and develop efficient broadband spintronic terahertz emitters [19–22].

Several recent works have highlighted the importance of the orbit degree of freedom in condensed matter physics and kicked off the emergent research field of orbitronics [23]. Orbitronics exploits the transport of orbital angular momentum through materials by orbit currents which can be exploited as an information carrier in solid-state devices. As for spin and charge current in SHE, it has been theoretically predicted and experimentally shown that a charge current $j_C$ can be converted into an orbit current $j_L$ via the orbital Hall effect (OHE) or the orbital Rashba-Edelstein effect (OREE) [24–31].

The orbit current cannot exert a torque directly on magnetization due to the lack of direct coupling between the orbit current and the magnetization. However, the spin-orbit coupling can convert the orbit current $j_L$ into spin current $j_S$, thereby generating torques on the magnetization. The conversion efficiency depends on the strength of spin-orbit coupling. Several studies have identified the spin torques originating from OHE in heterostructures [29,30].

According to the Onsager reciprocal relation, the inverse effect of the OHE should convert an orbit current into a charge current. To date, few studies reported the conversion from orbit current to charge current. This is due to the lack of a reliable orbit current source and the disturbance of the omnipresent inverse spin Hall effect. In this paper, we provide strong evidence for conversion from a current orbital $j_L$ into a charge current through free-space terahertz emission. The new effect, the orbital counterpart of ISHE, is named the inverse orbital Hall effect (IOHE) [23]. Thanks to the induced charge current, the IOHE provides a feasible approach for the electrical characterization of the orbit current.

In this study, magnetic multilayer samples were prepared by high-vacuum magnetron sputtering on a glass substrate. Unless otherwise specified, all the samples are seeded on a MgO buffer and protected with a capping layer of a MgO(2 nm)/Ta (2 nm) bilayer. The terahertz (THz) emission from those samples was studied at room temperature using a homemade THz time-domain spectroscopy. Femtosecond pulses of 35-fs pulse duration induce ultrafast charge current in our samples, which emits terahertz electromagnetic waves in the free space. The generation of the ultrafast charge current in magnetic multilayers can be studied by sampling the THz waves in free space.

We begin by comparing THz waveforms from MgO(3 nm)/NM (4 nm)/CoFeB (10 nm)/MgO/Ta and MgO(3 nm)/NM (4 nm)/Ni (10 nm)/MgO/Ta, where NM = Ta, Pt, or Cu (Figure 1). The NM/CoFeB samples followed the typical behavior of spintronic terahertz emission driven by spin-charge conversion via ISHE. The ultrafast charge current induced by the spin current is given as $j_C = \gamma_S j_S$, where $j_S$ refers to the spin current, and $\gamma_S$ is the spin-to-charge efficiency of the NM layer. For a given sample (e.g., CoFeB/Pt), the waveform polarity is reversed when the sample is flipped from front excitation to back excitation. Moreover, the waveform polarities of NM/CoFeB, dictated by the sign of the SHE of NM, are opposite for Ta relative to Pt, as expected from their opposite SHE [4, 5]. In agreement with the very small spin-orbit interaction and negligible ISHE in Cu, the sample CoFeB/Cu shows a small signal very similar to the signal obtained with a CoFeB monolayer without any evidence of additional contribution from Cu. The intrinsic signals in CoFeB monolayers can be ascribed to the anomalous Hall effect (AHE) of the CoFeB layer [32,33], as will be discussed later.

In contrast, the waveforms of NM(4 nm)/Ni(10 nm) showed reversed polarities with respect to Ni monolayer and the same polarities for all three samples, regardless of the sign of spin Hall angle of the NM layers (Figure 1b). Due to the opposite spin Hall angle of Pt and Ta, spin-charge conversion via ISHE should give opposite polarities for Pt/Ni and Ta/Ni. In addition, as Cu(4

nm)/Ni(10 nm) shows a waveform opposite to the intrinsic one of the Ni monolayer, this reversal reveals a significant additional contribution from Cu in spite of its minimal spin Hall angle. Our results, the same polarity with Pt and Ta and considerable contribution from the addition of Cu, show that ISHE is not sufficient to explain the THz waveforms for Ni-based bilayers.

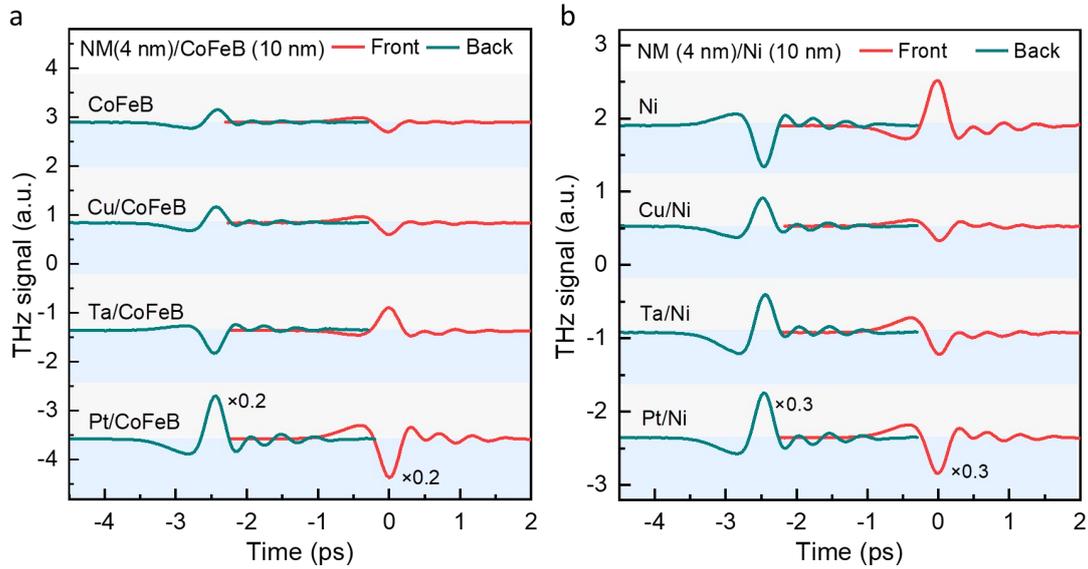

Figure 1 A comparison between THz signals from NM/Ni and NM/CoFeB heterostructures. a) THz waveforms emitted from CoFeB (10 nm), Cu (4 nm)/CoFeB (10 nm), Ta (4 nm)/CoFeB (10 nm), and Pt (4 nm)/CoFeB (10 nm) measured with front and back sample excitations. b) THz waveforms emitted from Ni (10 nm), Cu (4 nm)/Ni (10 nm), Ta (4 nm)/Ni (10 nm), and Pt (4 nm)/ Ni (10 nm) measured with front and back sample excitations. The above experiments were carried out under the same experimental conditions with an in-plane magnetic field of 1000Oe.

Since the results of NM/Ni cannot be fully explained with the spin-to-charge conversion by ISHE, we examine the possible underlying mechanisms that enable light-induced THz emission in the Ni-based samples. To date, several mechanisms are known to induce THz emission in magnetic thin films, including magnetic dipole emission, THz emission driven by AHE, and THz emission driven by spin-to-charge conversion or orbit-to-charge conversion.

First, magnetic dipole emission in Ni, first reported by Beaurepaire et al., scales with saturation magnetization. In magnetic thin films, magnetic dipole emission is very weak. Experimentally, magnetic dipole emission can be identified by flipping the samples. If magnetic dipole emission were the driving mechanism, we would expect THz waveforms of the same polarity for front and back excitation and similar THz amplitude for different NM materials, which is not supported by our results (Figure 1). Therefore, the THz emission from Ni-based samples is not driven by the magnetic dipole emission.

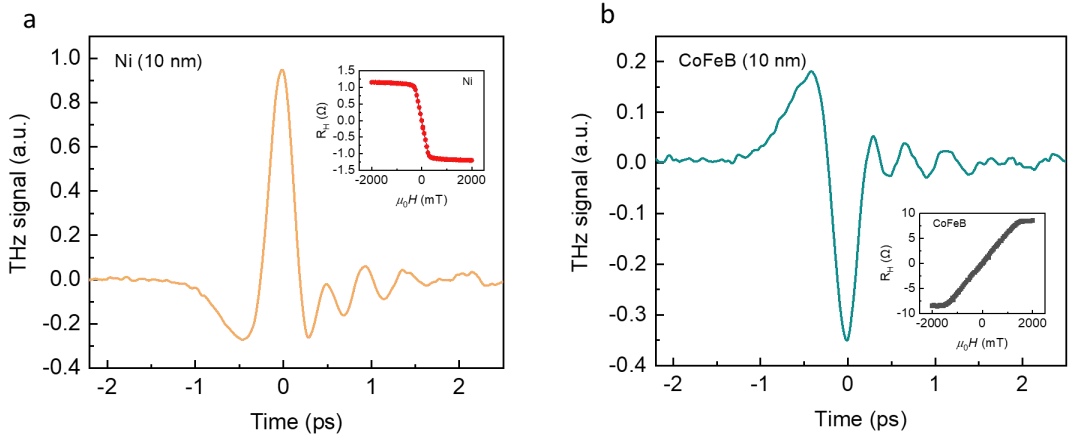

Figure 2 a) THz waveforms emitted from the Ni (10 nm) monolayer. The inset shows the Hall resistance $R_H$ as a function of the out-of-plane magnetic fields H for the Ni sample. b) THz waveforms emitted from the CoFeB (10 nm) monolayer. The inset shows the Hall resistance $R_H$ as a function of the out-of-plane magnetic fields H for the CoFeB sample.

The second possible contribution to THz emission comes from the AHE of the ferromagnetic metal, as proposed by Zhang et al. to explain the THz emission from a single ferromagnetic layer of FeMnPt [32]. In a metallic ferromagnetic layer, light-induced hot electrons are differently reflected by asymmetric interfaces, thus generating a transient charge current from one interface to the other. The deflection of this charge current by AHE induces in-plane charge currents current $j_C$ and spin currents $j_S$ flowing in the in-plane direction perpendicular to the magnetization. The charge current can be written as $j_C = \gamma_{AHE} j_l$ where the deflection coefficient $\gamma_{AHE}$ is equal to the AHE angle of the ferromagnetic layer. The resulting contribution to THZ emission is proportional to $\gamma_{AHE}$ [32,34]. As illustrated in the insets of Fig. 2, the AHE angles of Ni (~ -0.004) [35] and CoFeB [36] are opposite, and hence THz waveforms of opposite polarities are expected for Ni and CoFeB monolayers. Based on the results in Fig.1 and 2, the intrinsic emissions by Ni and CoFeB monolayers are predominantly related to the AHE.

The spin-to-charge conversion is the third mechanism for THz emission in NM/FM bilayers.

For the NM/CoFeB bilayers with NM = Cu, Ta, or Pt, the THz signal for Cu/CoFeB is negligibly different from that of a CoFeB monolayer in agreement with the very small SHE of Cu, while Ta and Pt add significant contributions of opposite signs for Ta and Pt, as expected for spin-to-charge conversion by the opposite ISHE of Ta and Pt [4, 5].

For the NM/Ni bilayers, the addition of Ta, Pt, or, as well, Cu leads to significant emissions of the same polarity, which is reversed with respect to the polarity of the AHE-induced emission by Ni monolayers without NM. Since the spin Hall angle is very small for Cu, this reversal with respect to the AHE-induced signal in the Ni monolayer cannot be explained by ISHE. The same polarity of the signals for Ta and Pt is also in contradiction to the conversion of spin current to charge current since Ta and Pt have opposite SHEs. As in very recent THz results with Ni [37], our results can be explained by a significant light-induced generation of orbit currents in Ni, as illustrated by Figure 3a. Among the contributions to THz emission, the charge currents generated by orbit-to-charge conversion of these orbit currents in the NM layer dominates over both the AHE intrinsic signal of Ni and the spin-to-charge ISHE conversion of the light-induced spin currents. Quantitatively we can write for the resulting charge current: $\boldsymbol{j_C} = \gamma_{AHE}\boldsymbol{j}_l + \gamma_L\boldsymbol{j_L} + \gamma_S\boldsymbol{j_S}$ where $\gamma_L$ and $\gamma_S$ are the orbit-to-charge and spin-to-charge conversion coefficients by IOHE and ISHE. Our results, with the polarities for Cu, Ta, and Pt reversed with respect to Ni monolayer and the polarities of Ta and Pt aligned in the same direction, show the predominance of the orbit term on the AHE and ISHE terms to align all the polarities finally. They also confirm that the sign of $\gamma_L$ is the same for Pt, Cu, and Ta, in agreement with theoretical predictions of the same sign of the IOHE in large series of nonmagnetic metals [28,38]. As we have seen, the results with CoFeB present the different signs expected for ISHE and the absence of a significant signal for Cu/CoFeB. They also indicate that CoFeB is much less efficient for the production of light-induced orbit currents.

For an additional characterization of the orbital contribution, we measured the THz waveforms from Ni(5 nm)/Pt and CoFeB(5 nm)/Pt, with the thickness of Pt varied between 1 and 6 nm (Figure 3b and c). The peak THz amplitude divided by saturation magnetization as a function of the thickness of Pt is shown in Figure 3d. It turns out that Ni/Pt is much more efficient in generating THz emission than CFB/Pt, which reflects both the efficient production of orbit current by Ni and the addition of IOHE and ISHE contributions of the same sign for Pt. The maximum emission is obtained at about 2nm of Pt, which suggests a propagation length of spin and orbit currents in this range for the ballistic electrons injected into Pt, similar to or slightly smaller than what is generally found for Fermi energy electrons in Pt [4]. Another information comes from the negligible shift in time of the waveforms as a function of the Pt thickness. Up to now, we have considered only IOHE and ISHE, whereas, in principle, interfaces of the NM layer could introduce additional contributions to the conversions by Inverse Orbital Rashba Edelstein Effect (IOREE) and Inverse Spin Rashba Edelstein Effects (ISREE). In similar bilayers, it has been shown that interfacial conversions introduce time shifts in the waveforms of THz emission [37]. The absence of such shifts in Fig.3b-c indicates that interfacial conversions are not significant in our experiments with respect to conversions by IOHE and ISHE.

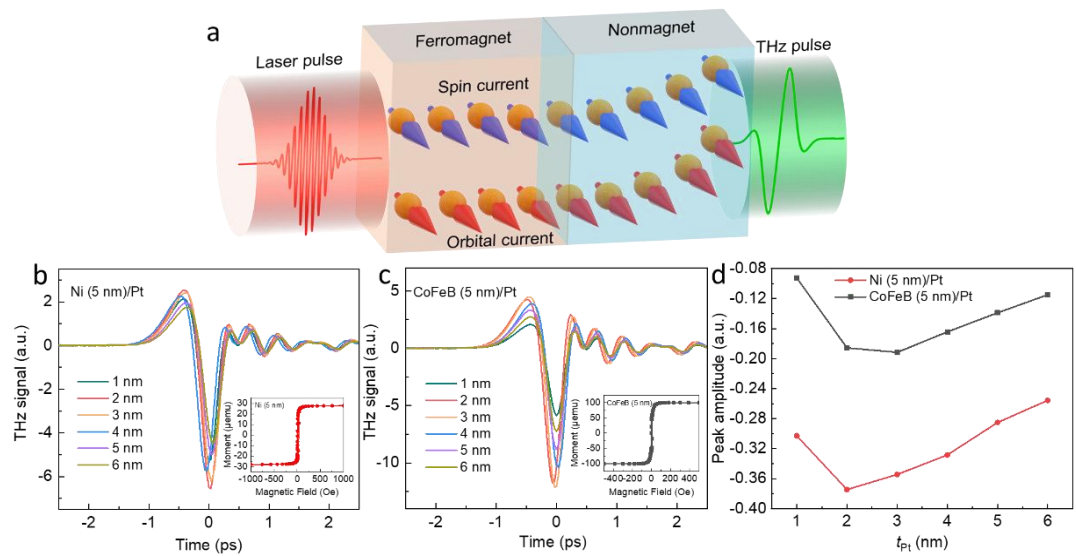

Figure 3 a) Conceptional diagram of the THz emission showing the conversion from spin/orbit currents into charge currents in the nonmagnetic layer. b) THz waveforms emitted from the Ni (5 nm)/Pt bilayers with different Pt thicknesses. The inset shows the hysteresis loops of Ni (5 nm). c) THz waveforms emitted from the CoFeB (5 nm)/Pt bilayers with different Pt thicknesses. The inset shows the vibrating sample magnetometer hysteresis loops of CoFeB (5 nm). d) The terahertz peak amplitude of Ni (5 nm)/Pt and CoFeB (5 nm)/Pt bilayers are divided by saturation magnetization of Ni or CoFeB and plotted against the Pt thickness.

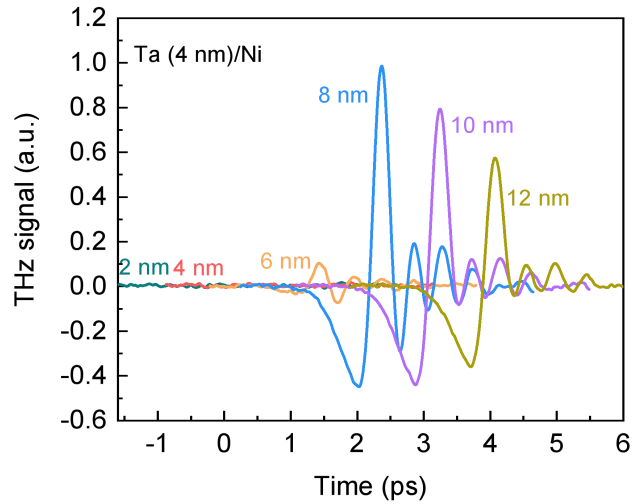

Figure 4 THz emission signals generated from Ta (4 nm)/Ni bilayers with different Ni thicknesses. The waveforms are shifted horizontally for clarity.

To further understand the conversion from orbit currents into charge currents in Ni-based bilayers, we analyzed the dependence on the Ni thickness for the Ta(4 nm)/Ni bilayer in Figure 4. The THz signal appeared above a threshold thickness, approximately 6 nm for Ni. This effective threshold Ni thickness corresponds to the minimum thickness to have a Curie temperature high enough to maintain ferromagnetic order along the applied field during laser excitation. Therefore, in agreement with the need for broken time-reversal symmetry, a well-established magnetization is necessary for creating such orbit currents by light (at least for non-circularly polarized light).

The usual mechanisms of generating orbit currents are from the charge-to-orbit conversions via OHE/OREE or spin-to-orbit conversions via spin-orbit coupling. The most striking result of our work is the light-induced generation of orbit current in Ni. We could not find any theoretical work predicting the production of orbit current in the situation of high-energy excitation. For electrons close to the Fermi level and for the interpretation of torques induced by orbit currents, several theoretical works have been devoted to the calculation of the coefficient of the conversions between orbit and spin current. For high-energy electrons, a first tentative scenario might be the production of orbit current from the light-induced spin current. For low energy electrons, according to the coefficients calculated for the conversion between orbit and spin current in Fig.2b in [29], we can see that the orbit-to-spin conversion is definitely more efficient in Ni compared to other 3d metals or alloys, which shows some similarity with our results. In another scenario, and more probably, the light-induced production of orbit current would not go via a conversion from spin current but directly from the excitation of orbital states. Further theoretical works are needed to study this type of mechanism. From a more general point of view, the specific character of Ni for the production of orbit current is also confirmed by other types of experiments. In the Spin Torque Ferromagnetic Resonance (ST-FMR) presented by Lee *et al*. [29], Ni is described as "abnormal" because CoFeB/Ta and Ni/Ta show opposite voltages generated in Ta at the resonance of the ferromagnet. Similarly, in our experiments (Fig.1), opposite THz emissions are generated by laser pulse respectively on CoFeB/Ta and Ni/Ta. This striking similarity between spin-torque ferromagnetic resonance and light-induced THz emission supports the idea that orbit currents can be generated not only by conversion from charge or spin currents but also by different types of excitations, including excitation by light and ferromagnetic resonance (FMR). Recently, the existence of orbit current generated by FMR and converted into voltage by orbital Rashba-Edelstein effect was also proposed in CFB/NM/Oxide systems [39].

In summary, orbitronics is an emerging and promising field of research in which our work introduces a new experimental method based on the production of light-induced orbit current and their exploitation for terahertz emission from NM/Ni bilayers with NM = Cu, Ta, or Pt. We find terahertz emission of the same polarity with Cu, Ta, and Pt as NM in NM/Ni bilayers despite the opposite SHE of Ta with respect to Pt. This common polarity is opposite to the polarity of the AHE-induced emission by a Ni monolayer. In addition, the same polarity for Ta and Pt cannot be explained by conversion by the opposite ISHE of Ta and Pt. Ni appears to be an efficient emitter of orbit current, and the terahertz emission generated by IOHE (orbit-to-charge conversion) in NM dominates the emission by ISHE (spin-to-charge conversion) and the emission generated by the AHE of the magnetic layer. Our results reveal how ballistic orbit currents can be efficiently induced by light and converted to charge current by IOHE for terahertz emission, opening a new route for developing future devices using the orbital degree of freedom.

# References

[1] J. Sinova, S. O. Valenzuela, J. Wunderlich, C. H. Back, and T. Jungwirth, *Spin Hall Effects*, Rev. Mod. Phys. **87**, 1213 (2015).

[2] A. Soumyanarayanan, N. Reyren, A. Fert, and C. Panagopoulos, *Emergent Phenomena Induced by Spin–Orbit Coupling at Surfaces and Interfaces*, Nature **539**, 7630 (2016).

[3] K. Ando, S. Takahashi, K. Harii, K. Sasage, J. Ieda, S. Maekawa, and E. Saitoh, *Electric Manipulation of Spin Relaxation Using the Spin Hall Effect*, Phys. Rev. Lett. **101**, 036601 (2008).

[4] A. Hoffmann, *Spin Hall Effects in Metals*, IEEE Trans. Magn. **49**, 5172 (2013).

[5] L. Liu, C. F. Pai, Y. Li, H. W. Tseng, D. C. Ralph, and R. A. Buhrman, *Spin-Torque Switching with the Giant Spin Hall Effect of Tantalum*, Science **336**, 555 (2012).

[6] I. M. Miron, K. Garello, G. Gaudin, P. J. Zermatten, M. V. Costache, S. Auffret, S. Bandiera, B. Rodmacq, A. Schuhl, and P. Gambardella, *Perpendicular Switching of a Single Ferromagnetic Layer Induced by In-Plane Current Injection*, Nature **476**, 189 (2011).

[7] V. M. Edelstein, *Spin Polarization of Conduction Electrons Induced by Electric Current in Two-Dimensional Asymmetric Electron Systems*, Solid State Commun. **73**, 233 (1990).

[8] Y. Fan et al., *Magnetization Switching through Giant Spin–Orbit Torque in a Magnetically Doped Topological Insulator Heterostructure*, Nat. Mater. **13**, 7 (2014).

[9] A. Manchon, H. C. Koo, J. Nitta, S. M. Frolov, and R. A. Duine, *New Perspectives for Rashba Spin–Orbit Coupling*, Nat. Mater. **14**, 871 (2015).

[10] A. R. Mellnik et al., *Spin-Transfer Torque Generated by a Topological Insulator*, Nature **511**, 449 (2014).

[11] A. Manchon, J. Železný, I. M. Miron, T. Jungwirth, J. Sinova, A. Thiaville, K. Garello, and P. Gambardella, *Current-Induced Spin-Orbit Torques in Ferromagnetic and Antiferromagnetic Systems*, Rev. Mod. Phys. **91**, 035004 (2019).

[12] P. Gambardella and I. M. Miron, *Current-Induced Spin–Orbit Torques*, Philos. Trans. R. Soc. Math. Phys. Eng. Sci. **369**, 3175 (2011).

[13] E. Saitoh, M. Ueda, H. Miyajima, and G. Tatara, *Conversion of Spin Current into Charge Current at Room Temperature: Inverse Spin-Hall Effect*, Appl. Phys. Lett. **88**, 182509 (2006).

[14] K. Shen, G. Vignale, and R. Raimondi, *Microscopic Theory of the Inverse Edelstein Effect*, Phys. Rev. Lett. **112**, 096601 (2014).

[15] K. Uchida, S. Takahashi, K. Harii, J. Ieda, W. Koshibae, K. Ando, S. Maekawa, and E. Saitoh, *Observation of the Spin Seebeck Effect*, Nature **455**, 778 (2008).

[16] O. Mosendz, J. E. Pearson, F. Y. Fradin, G. E. W. Bauer, S. D. Bader, and A. Hoffmann, *Quantifying Spin Hall Angles from Spin Pumping: Experiments and Theory*, Phys. Rev. Lett. **104**, 046601 (2010).

[17] J.-C. Rojas-Sánchez, N. Reyren, P. Laczkowski, W. Savero, J.-P. Attané, C. Deranlot, M. Jamet, J.-M. George, L. Vila, and H. Jaffrès, *Spin Pumping and Inverse Spin Hall Effect in Platinum: The Essential Role of Spin-Memory Loss at Metallic Interfaces*, Phys. Rev. Lett. **112**, 106602 (2014).

[18] J. C. Rojas-Sánchez et al., *Spin to Charge Conversion at Room Temperature by Spin Pumping into a New Type of Topological Insulator: $\alpha$-Sn Films*, Phys. Rev. Lett. **116**, 096602 (2016).

[19] M. B. Jungfleisch, Q. Zhang, W. Zhang, J. E. Pearson, R. D. Schaller, H. Wen, and A. Hoffmann, *Control of Terahertz Emission by Ultrafast Spin-Charge Current Conversion at Rashba Interfaces*, Phys. Rev. Lett. **120**, 207207 (2018).

[20] T. Seifert et al., *Efficient Metallic Spintronic Emitters of Ultrabroadband Terahertz Radiation*, Nat. Photonics **10**, 483 (2016).

[21] Y. Wu, M. Elyasi, X. Qiu, M. Chen, Y. Liu, L. Ke, and H. Yang, *High-Performance THz Emitters Based on Ferromagnetic/Nonmagnetic Heterostructures*, Adv. Mater. **29**, 1603031 (2017).

[22] C. Zhou et al., *Broadband Terahertz Generation via the Interface Inverse Rashba-Edelstein Effect*, Phys. Rev. Lett. **121**, 086801 (2018).

[23] D. Go, D. Jo, H.-W. Lee, M. Kläui, and Y. Mokrousov, *Orbitronics: Orbital Currents in Solids*, EPL Europhys. Lett. **135**, 37001 (2021).

[24] S. Ding et al., *Harnessing Orbital-to-Spin Conversion of Interfacial Orbital Currents for Efficient Spin-Orbit Torques*, Phys. Rev. Lett. **125**, 177201 (2020).

[25] D. Go and H.-W. Lee, *Orbital Torque: Torque Generation by Orbital Current Injection*, Phys. Rev. Res. **2**, 013177 (2020).

[26] D. Go, D. Jo, C. Kim, and H.-W. Lee, *Intrinsic Spin and Orbital Hall Effects from Orbital Texture*, Phys. Rev. Lett. **121**, 086602 (2018).

[27] D. Go, D. Jo, T. Gao, K. Ando, S. Blügel, H.-W. Lee, and Y. Mokrousov, *Orbital Rashba Effect in a Surface-Oxidized Cu Film*, Phys. Rev. B **103**, L121113 (2021).

[28] H. Kontani, T. Tanaka, D. S. Hirashima, K. Yamada, and J. Inoue, *Giant Orbital Hall Effect in Transition Metals: Origin of Large Spin and Anomalous Hall Effects*, Phys. Rev. Lett. **102**, 016601 (2009).

[29] D. Lee et al., *Orbital Torque in Magnetic Bilayers*, Nat. Commun. **12**, 1 (2021).

[30] S. Lee et al., *Efficient Conversion of Orbital Hall Current to Spin Current for Spin-Orbit Torque Switching*, Commun. Phys. **4**, 1 (2021).

[31] T. Tanaka, H. Kontani, M. Naito, T. Naito, D. S. Hirashima, K. Yamada, and J. Inoue, *Intrinsic Spin Hall Effect and Orbital Hall Effect in 4 d and 5 d Transition Metals*, Phys. Rev. B **77**, 165117 (2008).

[32] Q. Zhang, Z. Luo, H. Li, Y. Yang, X. Zhang, and Y. Wu, *Terahertz Emission from Anomalous Hall Effect in a Single-Layer Ferromagnet*, Phys. Rev. Appl. **12**, 054027 (2019).

[33] Y. Liu, H. Cheng, Y. Xu, P. Vallobra, S. Eimer, X. Zhang, X. Wu, T. Nie, and W. ZHAO, *Separation of Emission Mechanisms in Spintronic Terahertz Emitters*, Phys. Rev. B **104**, 064419 (2021).

[34] Y. Yang, Z. Luo, H. Wu, Y. Xu, R.-W. Li, S. J. Pennycook, S. Zhang, and Y. Wu, *Anomalous Hall Magnetoresistance in a Ferromagnet*, Nat. Commun. **9**, 2255 (2018).

[35] T. Miyasato, N. Abe, T. Fujii, A. Asamitsu, S. Onoda, Y. Onose, N. Nagaosa, and Y. Tokura, *Crossover Behavior of the Anomalous Hall Effect and Anomalous Nernst Effect in Itinerant Ferromagnets*, Phys. Rev. Lett. **99**, 086602 (2007).

[36] S. B. Wu, X. F. Yang, S. Chen, and T. Zhu, *Scaling of the Anomalous Hall Effect in Perpendicular CoFeB/Pt Multilayers*, J. Appl. Phys. **113**, 17C119 (2013).

[37] T. S. Seifert, D. Go, H. Hayashi, R. Rouzegar, F. Freimuth, K. Ando, Y. Mokrousov, and T. Kampfrath, *Time-Domain Observation of Ballistic Orbital-Angular-Momentum Currents with Giant Relaxation Length in Tungsten*, arXiv:2301.00747.

[38] L. Salemi and P. M. Oppeneer, *First-Principles Theory of Intrinsic Spin and Orbital Hall and Nernst Effects in Metallic Monoatomic Crystals*, arXiv:2203.17037.

[39] E. Santos, J. E. Abrão, D. Go, L. K. de Assis, J. B. S. Mendes, and A. Azevedo, *Inverse Orbital Torque via Spin-Orbital Entangled States*.

## Acknowledgements

The authors gratefully acknowledge the National Key Research and Development Program of China (No. 2022YFB4400200), National Natural Science Foundation of China (No. 92164206, No. 11904016, No. 52261145694 and No. 52121001), Beihang Hefei Innovation Research Institute Project (BHKX-19-01, BHKX-19-02). All the authors sincerely thanks Hefei Truth Equipment Co., Ltd for the help on film deposition. AF thanks Dr. H. Jaffres for fruitful discussions. This work was supported by the Tencent Foundation through the XPLORER PRIZE